\documentclass[preprint,superscriptaddress,longbibliography]{revtex4-1}
\usepackage{braket}
\usepackage{amsmath}
\usepackage{bm}
\usepackage{graphicx}
\usepackage[
	colorlinks=true,
	linkcolor=blue,
	citecolor=blue,
	urlcolor=blue,
	breaklinks
]{hyperref}

\begin{document}

\title{Experimental and Theoretical Study of Thermodynamic Effects in a Quantum Annealer}

\author{Tadashi Kadowaki}
\affiliation{DENSO CORPORATION, Nihonbashi, Chuo-ku, Tokyo 103-6015, Japan}
\author{Masayuki Ohzeki}
\affiliation{Graduate School of Information Science, Tohoku University, Sendai 980-8579, Japan}
\begin{abstract}
Quantum devices are affected by intrinsic and environmental noises.
An in-depth characterization of noise effects is essential for exploiting noisy quantum computing.
To this end, we studied the energy dissipative behavior of a quantum annealer via experiments and numerical simulations.
Our investigation adopts a recently proposed technique that interpolates between pure quantum dynamics and pure thermodynamics.
Experiments were conducted on a quantum annealer with an anneal pause function, which inserts a thermal relaxation period into the annealing schedule by pausing the transverse field, which is a source of quantum fluctuation.
After investigating the special Hamiltonian that characterizes the quantum thermodynamics of the system, we then observed enhancement of thermodynamic signature depending on the anneal pause parameter.
The time development of the state vector, observed in the open quantum simulation, provides rich information for investigating phenomena beyond energy-gap analysis.
We identified a special eigenstate bridges ground states far-separated in Hilbert space and the transfer probabilities from one ground state to another.
This finding can improve the sampling uniformity by reducing the sampling bias in finding the classical ground states in the quantum annealer.
Our study does not only characterize the open quantum phenomenon of the specific Hamiltonian but also demonstrates the usefulness of the method in investigating noisy quantum devices.
\end{abstract}
\maketitle

\section{Introduction}

Quantum mechanics is sufficiently powerful to explore all possible combinations of variables in quadratic unconstrained binary optimization (QUBO) problems.
To exploit this power, researchers have investigated quantum annealing (QA) in the Ising model \cite{Kadowaki1998,Kadowaki1998a,Brooke1999,Farhi2001,Santoro2002,Santoro2006,Das2008,Morita2008,Tanaka2017}.
In the framework of the Ising model, QUBO problems can be represented with spin-1/2 particles as the binary variables, whose arbitrary relations are encoded as $p$-body ($p \ge 2$) interactions and their local fields\cite{Billionnet1989,Biamonte2008,Lucas2014}.
Using a hardware implementation of QA developed by D-Wave Systems Inc., in which the spin-1/2 variables are superconducting flux qubits, we can evaluate algorithms for QUBO problems in real-world settings\cite{Johnson2011}.

Reducing the thermal fluctuations and other noise sources would lengthen the coherence time and improve the reliability of quantum information processing on qubits, but is challenging in practice.
At the same time, the performance to solve QUBO problems by QA can be improved by thermal relaxation after anticrossing, which recovers the ground state probability from excited states\cite{Amin2008,Venuti2017,Dickson2013}.
Analyzing open quantum systems from this perspective would provide new insights, not only for improving the coherence time and evaluating device limitations but also for understanding the synergy between quantum and thermal effects, thereby improving device performance.

Open quantum systems have been studied by perturbation approaches such as the Redfield and Lindblad equations\cite{Redfield1965,Kossakowski1972}, which handle the microscopic interactions between the system and the bath.
Alternatively, we have analyzed open quantum systems by interpolating between quantum dynamics and thermodynamics\cite{Kadowaki2018}.
In the interpolated dynamics (ID), the coupling strength between the system and its environment is parameterized by an interpolation ratio, negating the need for implementing the system--bath interactions.
This method can analyze the system under any parameter condition.
At one extreme of the interpolation, ID recovers the closed quantum dynamics; at the other extreme, ID provides the classical thermodynamics.
Between these two extremes, the dynamics are mixed.
When the interpolation ratio is small (i.e., the coupling is weak), the system is dominated by quantum dynamics and is perturbed by thermodynamics.
The dynamics of a two-level system derived by this method generalize the optical Bloch equations\cite{Bloch1946,Arecchi1965}.
At the weak-coupling and asymptotic limits, the Bloch equations and our model equations become equivalent.

% add
Differential equations of the dynamics are derived from a continuous-time limit of repeated mixing process of the two solutions from quantum dynamics and thermodynamics.
This methodology is similar to stochastic Shr\"odinger equations (SSE) by taking a continuous-time limit of repeated quantum measurements of the environment\cite{Pellegrini2008,Pellegrini2010}.
In SSE, a system interacts with ``a copy'' of the environment for a short period, and quantum measurement of the environment is conducted and repeats the same process with ``other copies'' of the environment.
The measurement of the environment affects the system indirectly by a projection as they have interacted before the measurement.
Note that the solution is non-deterministic due to quantum jumps (stochastic projections introduced by quantum measurements).
Solutions of the equations are called quantum trajectories.
Continuous-time limit of the equations results in differential equations, and depending on the measurement type, either jump-type or diffusive-type differential equations are obtained.

\begin{figure}[thb]
	\center
    \includegraphics[width=85mm]{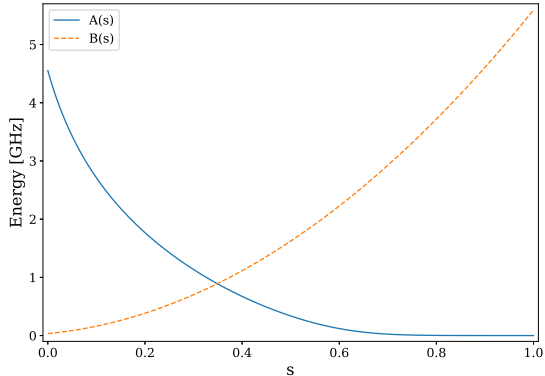}%
    \caption{
        \label{fig_schedule_curves}
        (Color online) The scheduling parameters $A(s)$ and $B(s)$ as a function of $s$ in the D-Wave 2000Q.
        For a standard annealing schedule we choose $s(t) = t/\tau$, where $\tau$ is a total annealing period.
        In general, $s(t)$ can be an arbitrary piecewise linear function in $[0,1]$ from 0 at the start to 1 at the end.
        }
\end{figure}

While the system is continuously interfered by the indirect continuous measurement in SSE, another continuous interference is adopted in ID.
Here, we briefly review ID, the method for open quantum systems\cite{Kadowaki2018}.
We construct ID from Schr\"odinger equation and classical master equation, which is determined by diagonal elements of the Hamiltonian in the Schro\"dinger equation.
Schr\"odinger equation in the natural unit ($\hbar=1$) is
\begin{equation}
 \label{eq_shrodinger}
  \frac{d}{dt} \ket{\psi(t)} = -i{\mathcal H}(t) \ket{\psi(t)} ,
\end{equation}
where the time-dependent Hamiltonian is defined by
\begin{equation}
  {\mathcal H}(t) = B(t) \ {\mathcal H}_c + A(t) \ {\mathcal H}_q .
\end{equation}
In QA, we adopt ${\mathcal H}_c$ is a problem Hamiltonian representing QUBO problems using diagonal operator $\sigma^z$,
\begin{equation}
  {\mathcal H}_c = - \sum_{(ij)} J_{ij} \sigma_i^z \sigma_j^z - \sum_i h_i \sigma_i^z ,
\end{equation}
and ${\mathcal H}_q$ is represented by off-diagonal operator $\sigma^x$,
\begin{equation}
  {\mathcal H}_q = - \sum_i \sigma_i^x .
\end{equation}
Time dependence of the Hamiltonian is determined through $A(t)$ and $B(t)$.
These scheduling parameters $A(t)$ and $B(t)$ change monotonically from 0 to 1 and from 1 to 0 respectively.
Therefore, the initial Hamiltonian is $\mathcal{H}_q$, which is the local transverse field, and the ground state is a superposition of all possible states.
The final Hamiltonian is $\mathcal{H}_c$ at the end of the dynamics, and the ground state realizes the solution of the QUBO problem.
The actual curves of the parameters in this study are depicted in Fig.~\ref{fig_schedule_curves}.
$A$ and $B$ are functions of not $t$ directly but indirectly through $s(t)$, where $0 \le s \le 1$.
We can design the shape of $s(t)$.

Master equation describes time-development of a probability distribution,
\begin{equation}
  \label{eq_master}
  \frac{d P_i(t)}{dt} = \sum_j {\mathcal L}_{ij} P_j(t) ,
\end{equation}
where $P_j(t)$ is the probability of the $j$-th state and ${\mathcal L}_{ij}$ is the transition rate matrix.
This matrix is characterized by the temperature $T$ and energy levels $\{E_i\}$ of ${\mathcal H}_c$,
\begin{equation}
  {\mathcal L}_{ij} =
  \begin{cases}
    \ \Big[ 1 + e^{\frac{E_i - E_j}{T}} \Big]^{-1} & (\mbox{single-spin flip}) \\
    \ - \sum_{k\neq i} {\cal L}_{ki} & (i=j) \\
    \ 0 & (\mbox{otherwise})
  \end{cases} ,
\end{equation}
where we use the natural unit ($k_B=1$).

Solutions for Schr\"odinger equation can be expressed as a linear combination of basis states with complex coefficients,
\begin{equation}
    \ket{\psi(t)} = \sum_i c_i(t) \ket{i} .
\end{equation}
We assume that both dynamics start from a compatible states, a trivial ground states of the given initial Hamiltonian, i.e. a superposition of all possible $z$-basis states for Schr\"odinger equation and uniform distribution for master equation,
\begin{align}
    c_i(0) & = 1 / \sqrt{N} , \\
    P_i(0) & = 1 / N ,
\end{align}
where $N$ is the number of all possible states.
This assumption satisfies the relation $|c_i(0)|^2 = P_i(0)$ for any $i$.

Similar to the short-time development by the total Hamiltonian in SSE, the two systems develop respectively for a short period.
Although SSE introduces the system-environment interaction in the total Hamiltonian, ID introduces not directly but indirectly as a result of the latter part of this methodology.
To interact the two systems indirectly, we construct a new state vector based on the solution of Schr\"odinger equation with the solution of master equation as a source of thermal interference.
The new state vector with a parameter $\alpha$, which controls the strength of the thermal interference, and an associated probability distribution are defined by
\begin{align}
    \label{eq_interploate_psi}
    \ket{\tilde{\psi}(t+dt)} & = \sum_i \tilde{c}_i(t+dt) \ket{i} \\
    & = \sum_i \sqrt{r_i(t+dt)} \frac{c_i(t+dt)}{|c_i(t+dt)|} \ket{i} , \\
%    \sum_i \sqrt{(1-\alpha)|c_i(t+dt)|^2 + \alpha P_i(t+dt)} \frac{c_i(t+dt)}{|c_i(t+dt)|} \ket{i} ,
    \tilde{P}_i(t+dt) & = r_i(t+dt) ,
\end{align}
%and
%\begin{equation}
%    \label{eq_interploate_p}
%    \tilde{P}_i(t+dt) = r_i(t+dt) ,
%    \tilde{P}_i(t+dt) = (1-\alpha)|c_i(t+dt)|^2 + \alpha P_i(t+dt) .
%\end{equation}
where $r_i(t) = (1-\alpha)|c_i(t)|^2 + \alpha P_i(t)$.
With these definitions, the same relation assumed in the initial state also satisfies
\begin{equation}
	|\tilde{c}_i(t+dt)|^2 = \tilde{P}_i(t+dt)
\end{equation}
for any $i$.
Through $r_i(t)$, the new state vector $\ket{\tilde{\psi}(t)}$ carries all information needed for ID.
As we interpolate solutions between Schr\"odinger and master equations parametrized by $\alpha$, the new state vector becomes a solution of Schr\"odinger equation in the case of $\alpha = 0$, and master equation in $\alpha = 1$.

We adopt $\ket{\tilde{\psi}(t+dt)}$ and $\{\tilde{P}_i(t+d)\}$ as a state vector and a probability distribution at $t+dt$ insted of $\ket{\psi(t+dt)}$ and $\{P_i(t+d)\}$ for further time-development by Schr\"odinger equation and master equation.
This artificial ``jump'' form $\ket{\psi}$ to $\ket{\tilde{\psi}}$ (and from $\{P_i\}$ to $\{\tilde{P}_i\}$) is correspond to the stochastic projection by measurement in SSE.
While SSE is non-deterministic, ID is deterministic.
Difference equations (discrete jump process) can be obtained by substituting $c_i(t+dt) = c_i(t) - i \sum_j \mathcal{H}_{ij} c_j(t) dt$ from Eq.~(\ref{eq_shrodinger}) and $P_i(t+dt) = P_i(t) + \sum_j \mathcal{L}_{ij} P_j(t) dt$ from Eq.~(\ref{eq_master}) for Eq.~(\ref{eq_interploate_psi}), and applying the assumption, i.e, replacing $\ket{\tilde{\psi}(t+dt)} \to \ket{\psi(t+dt)}$ and $P_i(t) \to |c_i(t)|^2$.
Then, we have
\begin{multline}
	\label{eq_new_state}
    \ket{\psi(t+dt)} = \\
    \sum_i \sqrt{(1-\alpha)\bigl|c_i(t) - i \sum_j \mathcal{H}(t)_{ij} c_j(t) dt\bigr|^2 + \alpha \Bigl(|c_i(t)|^2 + \sum_j \mathcal{L}_{ij} |c_j(t)|^2 dt \Bigr)} \\
    \times \frac{c_i(t) - i \sum_j \mathcal{H}(t)_{ij} c_j(t) dt}{|c_i(t) - i \sum_j \mathcal{H}(t)_{ij} c_j(t) dt|} \ket{i} .
\end{multline}
A series expansion for $dt$ of this equation is
\begin{equation}
    \ket{\psi(t+dt)} = \sum_i c_i(t) \ket{i} + \mathcal{F}\big(\{c_i(t)\}, \mathcal{H}(t), \mathcal{L}, \alpha\big) dt + O(dt^2) ,
\end{equation}
where $\mathcal{F}(\cdots)$ is the first order coefficient.
This can be written as a difference equation,
\begin{equation}
    \ket{\psi(t+dt)} - \ket{\psi(t)} = \mathcal{F}\big(\{c_i(t)\}, \mathcal{H}(t), \mathcal{L}, \alpha\big) dt + O(dt^2) .
\end{equation}
Continuous-time limit gives differential equations of ID,
\begin{equation}
	\label{eq_differential}
    \frac{d}{dt}\ket{\psi(t)} = \mathcal{F}\big(\{c_i(t)\}, \mathcal{H}(t), \mathcal{L}, \alpha\big) .
\end{equation}

SSE and ID can be expressed in a wave function form as well as a density matrix form, because the system keeps pure state\cite{Pellegrini2008,Pellegrini2010}.
We reported the density matrix representation of ID for a two-level system \cite{Kadowaki2018}, and a wave function representation is provided in Appendix A.
The differential equations (\ref{eq_tl_up}) and (\ref{eq_tl_down}) consist of terms from Shr\"odinger equation (the first and the second terms) and master equation (the third term) as well as additional non-linear terms in the right-hand side (RHS).
The thermal bath is included in the dynamics not explicitly but indirectly through master equation.
As a result, system--bath interaction is introduced and the interpolation parameter $\alpha$ controls the strength of energy dissipation.
Continuous interference of a wave function by replacing with an interpolated state in our proposed dynamics and quantum jumps by measurements in stochastic Schr\"odinger equation have similar roles in terms of introducing thermal (environmental) effects.
% end

\begin{figure}[thb]
    \center
    \includegraphics[width=30mm]{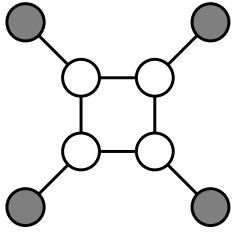}
    \caption{
        \label{fig_model}
        Eight-spin quantum-signature model.
        Each circle stands for a 1/2-spin, and spin-spin interactions are all ferromagnetic ($J_{ij}=1$).
        Local fields are $h_i=1$ on open circles (core spins) and $h_i=-1$ on closed circles (outer spins).
        If core spins are all up, any spin configurations on outer spins give the same energy level, and all of them are the ground states.
        In addition, all-spin-down configuration is the other ground state.
    }
\end{figure}

In the present paper, we conduct anneal pause experiments on the eight-spin quantum-signature model\cite{Boixo2013,Albash2015} depicted in Fig.~\ref{fig_model} using the D-Wave quantum annealer.
% This model comprises a ring of ``core spins'' with an ``outer spin'' attached to each core spin.
%All ferromagnetic interactions take the form $J_{ij} = 1$.
Here we consider the ferromagnetic case, and set all the interactions as $J_{ij} = 1$.
% The local fields of the core and outer spins are $h_i = 1$ and $-1$, respectively.
Due to the frustration caused by competition between the ferromagnetic interaction and the local field, the ground states degenerate into a cluster of 16 ground states with four up-spins in the core, and an isolated ground state with all eight core and outer spins being down, as follows:
\begin{align}
    C & = \{ \; \ket{\uparrow \uparrow \uparrow \uparrow \ \updownarrow \updownarrow \updownarrow \updownarrow} \, \} \;\;\; \text{(cluster of 16 states)} ,\\
    S & = \{ \; \ket{\downarrow \downarrow \downarrow \downarrow \ \downarrow \downarrow \downarrow \downarrow} \, \} \;\;\; \text{(isolated state)} ,
\end{align}
where four spins in the left(right) side represents core(outer) spins and $\updownarrow$ stands for both of up ($\uparrow$) and down ($\downarrow$) spin configurations.
Probabilities of the clustered and the isolated states are defined by
\begin{align}
	P_c & = \frac{1}{|C|} \sum_{j\in C} \big| \braket{\psi(t)|j} \big|^2 \\
    P_s & = \frac{1}{|S|} \sum_{j\in S} \big| \braket{\psi(t)|j} \big|^2 = \big| \braket{\psi(t)|S} \big|^2 .
\end{align}
Note that $P_c$ is the average of 16 states while $P_s$ is calculated from a single state.
The original study\cite{Boixo2013} found that the isolated state probability is suppressed in quantum dynamics and enhanced in thermodynamics.
A ratio $P_s / P_c$ is 1 at the thermodynamic equilibrium.
In other words, all ground states can be obtained with equal probability.
If the ratio is lower (higher) than 1, the isolated (clustered) state is suppressed and less probability. 
Therefore the ratio reflects the imbalance between two types of ground states.
In the case that the ratio is close to 1, all possible ground state configurations can be obtained efficiently, and the number of annealing trials can be minimized.
To this end, understanding the phenomena in the model helps to mitigate the suppression of the isolated state probability, and the ultimate goal is to figure out a general methodology to achieve $P_s / P_c \sim 1$.
This equal probability sampling is called ``fair sampling''\cite{Matsuda2009,Chancellor2017,Mandra2017,Konz2018}.

\begin{figure}[thb]
	\center
    \includegraphics[width=85mm]{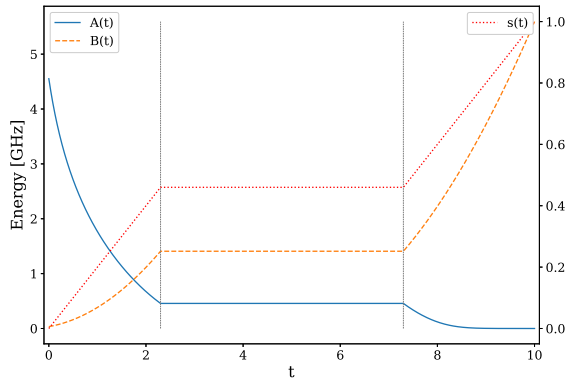}%
    \caption{
        \label{fig_anneal_pause_schedule}
        (Color online) Temporal changes in anneal schedule parameters $A(t)$ and $B(t)$ in anneal pause for $\tau_\mathrm{anneal}$ = $\tau_\mathrm{pause}$ = 5 $\mu$ and $s_\mathrm{pause}=$ 0.46.
        The parameters are determined through $s(t)$ according to the relationship in Fig.~\ref{fig_schedule_curves}.
        A pause period starts at $t$ = 2.3 and ends at $t$ = 7.3.
    }
\end{figure}

The anneal pause depicted in Fig.~\ref{fig_anneal_pause_schedule} is a special annealing schedule that pauses the change of the transverse field for a specified time, in other words, the waiting period.
During the anneal pause period, the quantum mechanics keep the system essentially in the same state as the Hamiltonian does not change.
Therefore, thermodynamics dominate despite the low temperature and sufficient isolation from the environment.
By changing the pausing level of the transverse field, we can change the Hamiltonian during the period, thereby controlling the thermodynamics.
Numerical simulations are expected to probe the detailed dynamics of the above-described open quantum system, beyond merely analyzing the energy-gap structure.
In particular, we investigate how the probability transfer from the clustered ground states to the isolated ground state mitigates the bias in the classical ground-state probabilities.

The remainder of this study is organized as follows:
Section~\ref{sec_experiments} describes the setup and results of the D-Wave experiments of the quantum-signature model.
Section~\ref{sec_simulations} provides the analytical expression of the differential equations of the model by expanding the interpolation method, performs the numerical simulations, and conducts data analysis of the dynamics.
Finally, Section~\ref{sec_summary} summarizes and discusses our results.

\section{D-Wave experiments}
\label{sec_experiments}

The anneal time $\tau_\mathrm{anneal}$ plus anneal pause period $\tau_\mathrm{pause}$ settings in the anneal pause experiments were varied as 1 + 1, 2 + 2, 5 + 5, and 10 + 10 $\mu$s.
Under each condition, 49 anneal pause parameters ($s_\mathrm{pause}=$ 0.02, 0.04, $\dots$, 0.98), normalized by the anneal time, were tested.
For example, $s_\mathrm{pause}=$ 0.46 during $\tau_\mathrm{anneal}$ = $\tau_\mathrm{pause}$ = 5 $\mu$s means that the annealing is paused at $t=$ 0.46 $\times$ 5 = 2.3 $\mu$s and lasts for 5 $\mu$s.
The remaining 2.7 $\mu$s is scheduled for the latter part of the annealing.
As a reference, we also performed continuous annealing without pause, designated $s_\mathrm{pause} = 0.0$ in the results.
The quantum-signature Hamiltonian can be mapped onto the unit Chimera graph of the D-Wave 2000Q device.
In this way, more than 250 copies are mapped and annealed simultaneously.
Each job submitted to the device comprised 1000 individual runs of the given Hamiltonian.
The runs were averaged to give the final results.
To remove the experimental biases, we repeated the measurements in two ways:
First, we mapped different patterns in the physical qubits to reduce the pattern bias.
For each unit Chimera graph, we selected ten mapping patterns from 144 (= 4! $\times$ 4! / 4) patterns implementing the quantum-signature Hamiltonian.
Second, we ran ten experiments, randomizing the job order by shuffling in each run, to cancel out the general biases depending on the order of experiments, such as the unexpected drift of the operation temperature, and other time-dependent biases.
By the above process, we generated $4\times(49 + 1)$$\times250$$\times1000$$\times10$$\times10 = $$ 5\times10^9$ data points.
The quantum processing time and communication time of the whole process was approximately one hour.

\begin{figure}[thb]
	\center
    \includegraphics[width=85mm]{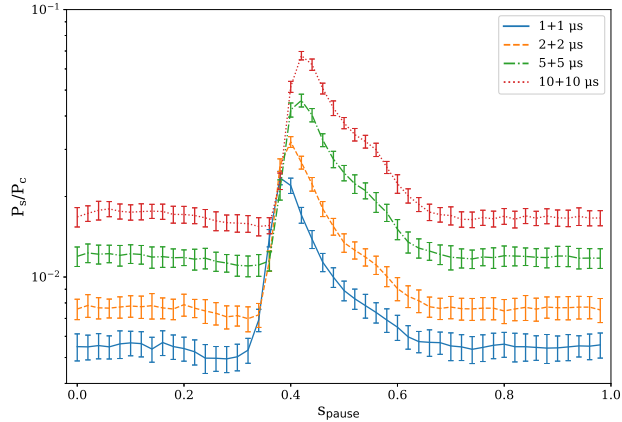}%
    \caption{
        \label{fig_experiments}
        (Color online) Anneal time plus pause period $\tau_\mathrm{anneal} + \tau_\mathrm{pause}$ and anneal pause level $s_\mathrm{pause}$ dependences of $P_s / P_c$ ratio.
        Error bars represent the standard deviations of 100 jobs for each data point.
        Among the 20,000 $P_s / P_c$ ratio data, we eliminated 33 ($\sim$0.17\%) data located outside 6 median absolute deviations (about 4 $\sigma$).
    }
\end{figure}

Figure~\ref{fig_experiments} plots the $P_s / P_c$ ratio versus anneal pause $s_\mathrm{pause}$ for various anneal time and pause periods $\tau_\mathrm{anneal} + \tau_\mathrm{pause}$.
During longer annealing time, more thermal effects accumulate through the system--environment coupling; hence, increasing the annealing time increases the ratio.
In addition, the ratio increased in the 0.3--0.7 range of $s_\mathrm{pause}$ under all four anneal-time conditions.
The thermal fluctuations drive the system more efficiently in this region than outside the region.
Although the isolated ground state is suppressed by quantum dynamics, it recovers during the anneal pause.
This mitigation of the imbalance between $P_s$ and $P_c$ is favorable for fair sampling.
The mechanism of the $s_\mathrm{pause}$-dependent increase of the ratio will be investigated in the next section.

\section{Numerical simulations}
\label{sec_simulations}

In our previous paper, we developed a method that represents open quantum systems by interpolating between quantum mechanics and thermodynamics.
Energy in the system dissipates not through an external thermal bath, but via the incorporated dynamics governed by the master equation.
We reported the dynamics of a single-spin system and larger systems by analytical and numerical expressions, respectively.
In the latter approach, we did not take a continuous-time limit to obtain final merged differential equations.
In the present paper, we expand our method to the analytical expression of larger systems.
The nonlinear differential equations of the quantum-signature Hamiltonian are symbolically obtained using Mathematica (version 11.3).

The RHS of Eq.~(\ref{eq_differential}) is obtained by the following steps:
(1) Construct the Schr\"odinger and master equations based on the given Hamiltonian in the $z$-basis.
(2) Using Eq.~(\ref{eq_interploate_psi}), calculate the total derivative of the interpolated wave function from the total derivatives of $c_i(t)$ and $P_i(t)$.
(3) Extract the first order coefficients of $dt$ by series expansion.
The coefficients of $dt$ are the RHSs of the differential equations of the system.
Appendix B provides the Mathematica code which generates C code of RHS of the dynamics, $\mathcal{F}(\cdots)$ in Eq.~(\ref{eq_differential}).
The final RHS expression constitutes $2^n$ complex variables: $n+1$ variables in each equation including own state and $n$ neighbor states connected by a single-spin flip.
The system size $n$ is eight in this study.
So that, we have 256 equations and each equation has nine terms in RHS.

\begin{figure}[htb]
    \center
    \includegraphics[width=85mm,clip]{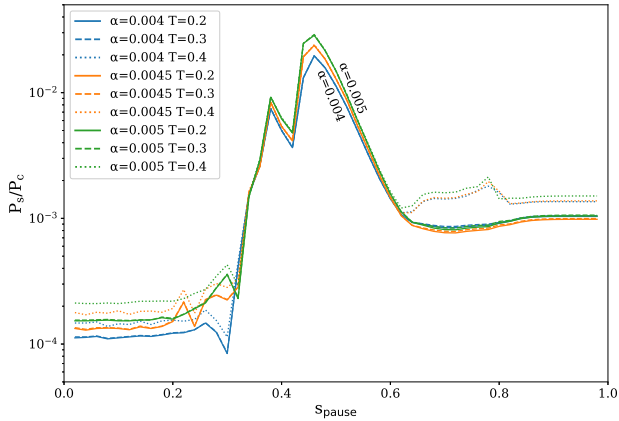}%
    \caption{
        \label{fig_prob_dep}
        (Color online) Effects of temperature ($T = $ 0.2, 0.3 and 0.4) and coupling strength ($\alpha = $ 0.004, 0.0045, and 0.005) on the $P_s / P_c$ ratio versus $s_\mathrm{pause}$ relation.
        %Large $T$ and $\alpha$ values enhance the thermal signature.
        Large $T$ value ($T=0.4$, dotted lines) enhances the thermal signature (larger $P_s / P_c$ ratio) for the low and high $s_\mathrm{pause}$ value ($s_\mathrm{pause} \lesssim 0.3$ and $s_\mathrm{pause} \gtrsim 0.6$), and large $\alpha$ value ($\alpha=0.005$, green solid line) does for the low and middle $s_\mathrm{pause}$ value ($s_\mathrm{pause} \lesssim 0.3$ and $0.4 \lesssim s_\mathrm{pause} \lesssim 0.5$).
    }
\end{figure}

As the energy scale of the Hamiltonian is normalized ($|J_{ij}| \sim 1$ and $|h_i| \sim 1$), the simulation parameters to be specified are the coupling constant (interpolation parameter) $\alpha$ and the temperature $T$.
To match the operation temperature of the D-Wave annealer (approximately 15 mK\cite{King2015}), we chose $T = 0.3$ GHz ($\sim$ 14.4 mK).
We conducted grid search of the coupling constant $\alpha$ to fit approximately the experimental peak ratio 0.0235 at $\tau_\mathrm{anneal}$ = $\tau_\mathrm{pause}$ = 1 $\mu$s to the experimental results.
This process was started from the grid search region [0.001, 0.04] followed by [0.003, 0.005].
Finally, we chose $\alpha$ = 0.0045 and the peak ratio is 0.0238, while the ratio is 0.0196 and 0.0290 for $\alpha$ = 0.004 and 0.005.
Although $s_\mathrm{pause}$ values for the peak of curves in the experiment and the numerical simulation does not match in our analysis, we prioritize to fit the peak ratio.
The effects of varying $T$ and $\alpha$ around the chosen values are shown in Fig.~\ref{fig_prob_dep}.

The 256 equations were solved by the ordinary differential equation (ODE) solver scipy.integrate.ode in SciPy.
The Adams method in the VODE solver was employed with nsteps = 200000 and atol = rtol = 1.136871e-13.
The ODE time step was selected as 0.01.
Smaller values of the time step (0.005, 0.002 and 0.001) did not change the results.
In each time step, the total length of the state vector was normalized to 1 to prevent accumulation of systematic drift.
The anneal scheduling parameters $A(s)$ and $B(s)$ of the D-Wave 2000Q applied to the numerical simulations were provided by D-Wave Systems Inc.

\begin{figure}[thb]
    \center
    \includegraphics[width=85mm]{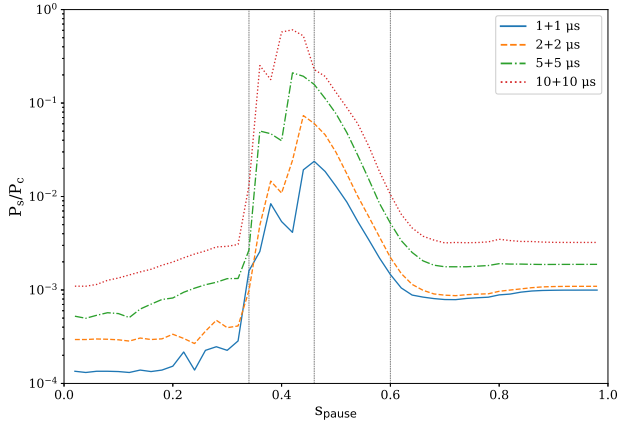}%
    \caption{
        \label{fig_simulations}
        (Color online) $P_s / P_c$ ratio versus anneal pause parameter $s_\mathrm{pause}$ for different anneal time plus pause periods $\tau_\mathrm{anneal}$ + $\tau_\mathrm{pause}$.
        Vertical lines mark the $s_\mathrm{pause}$ values of the peak in the curve for $\tau_\mathrm{anneal}$ + $\tau_\mathrm{pause}=$ 1 + 1 $\mu$s ($s_\mathrm{pause}$ = 0.46), and the low and high limits of the elevated-ratio range ($s_\mathrm{pause}$ = 0.32 and 0.60).
        The simulation parameters are $T$ = 0.3 and $\alpha$ = 0.0045.
    }
\end{figure}

\begin{figure}[thb]
    \center
    \includegraphics[width=85mm]{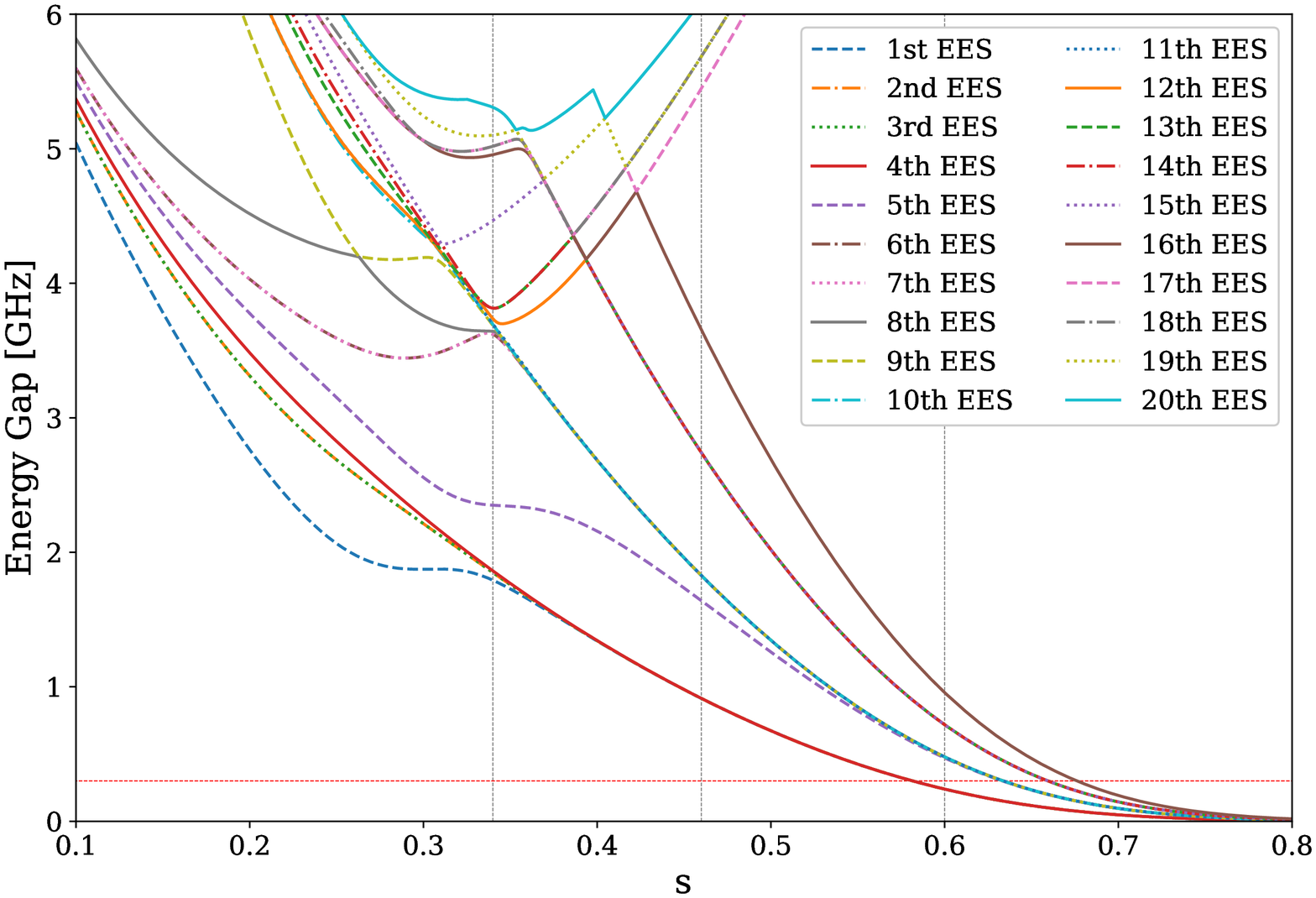}%
    \caption{
        \label{fig_energy_gap}
        (Color online) Energy gap (instantaneous energy levels from the ground eigenstate) versus anneal control parameter $s$, which determines the magnitude of the transverse field through $A(s)$.
        The lowest 20 excited eigenstates (EES) are drawn.
        Of 16 excited eigenstates converge to zero when $s = 1$ (17 ground states in total).
        Vertical dashed lines are drawn at $s$ = 0.34, 0.46, and 0.60.
        A horizontal dashed line in red represents the energy level of $T = 0.3$ GHz.
        The first, second, third, and fourth excited eigenstates seem to converge around $s =0.34$, and they go lower than 0.3 after $s = 0.6$
    }
\end{figure}

The simulation result (Fig.~\ref{fig_simulations}) reproduces the two fundamental characteristics in the experimental plot of Fig.~\ref{fig_experiments}; i.e., the anneal time and transverse field dependences of $P_s / P_c$.
The probability range was larger in the numerical simulation than in the experiments, and the detailed curve shapes slightly differed between the two sets of results.
To understand the phenomena that $P_s / P_c$ increases in a certain region (0.34--0.6 for $\tau_\mathrm{anneal}$ + $\tau_\mathrm{pause}=$ 1 + 1 $\mu$s), we investigated the energy-gap structure.
The energy gaps are calculated from the instantaneous eigenstates and eigenvalues.
If the system evolves adiabatically, it remains in the eigenstates with the lowest eigenvalue.
The instantaneous energy-gaps for the lowest 20 excited eigenstates as functions of $s$ are plotted in Fig.~\ref{fig_energy_gap}.
The instantaneous first excited eigenstate (blue dashed curve) around $s$ = 0.34 converges with three additional (second, third, and fourth) excited eigenstates.
The thermal energy level of the system (0.3 GHz, dotted horizontal line in red) exceeds the energy of the first excited eigenstate near $s$ = 0.60.
Above this value of $s$, change in the energy-gap structure is not dominant because the thermal fluctuation level is higher than the energy gap.
In other words, the system freezes out, and the probabilities do not change during the anneal pause period\cite{Amin2015}.

As the anneal control parameter $\tau_\mathrm{anneal}$ + $\tau_\mathrm{pause}$ increased, the simulated system exhibited more thermodynamic behavior, i.e. larger $P_s / P_c$ was observed as in Fig.~\ref{fig_simulations}.
Static analysis, such as energy-gap structure analysis, is of limited help in understanding such behavior.
Although additional static analyses can provide further information, they cannot inform beyond the dynamical analysis, as discussed later.

\begin{figure}[thb]
    \center
    (a) \includegraphics[width=85mm,clip]{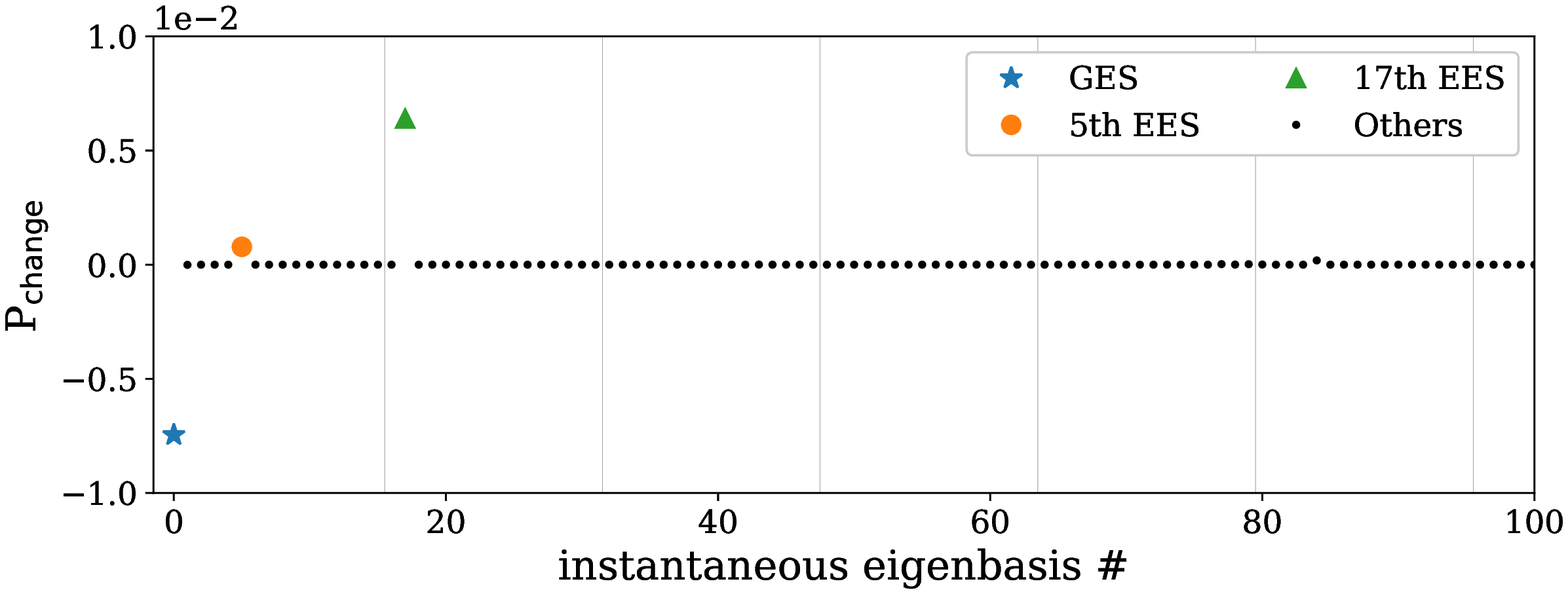} \\
    (b) \includegraphics[width=85mm,clip]{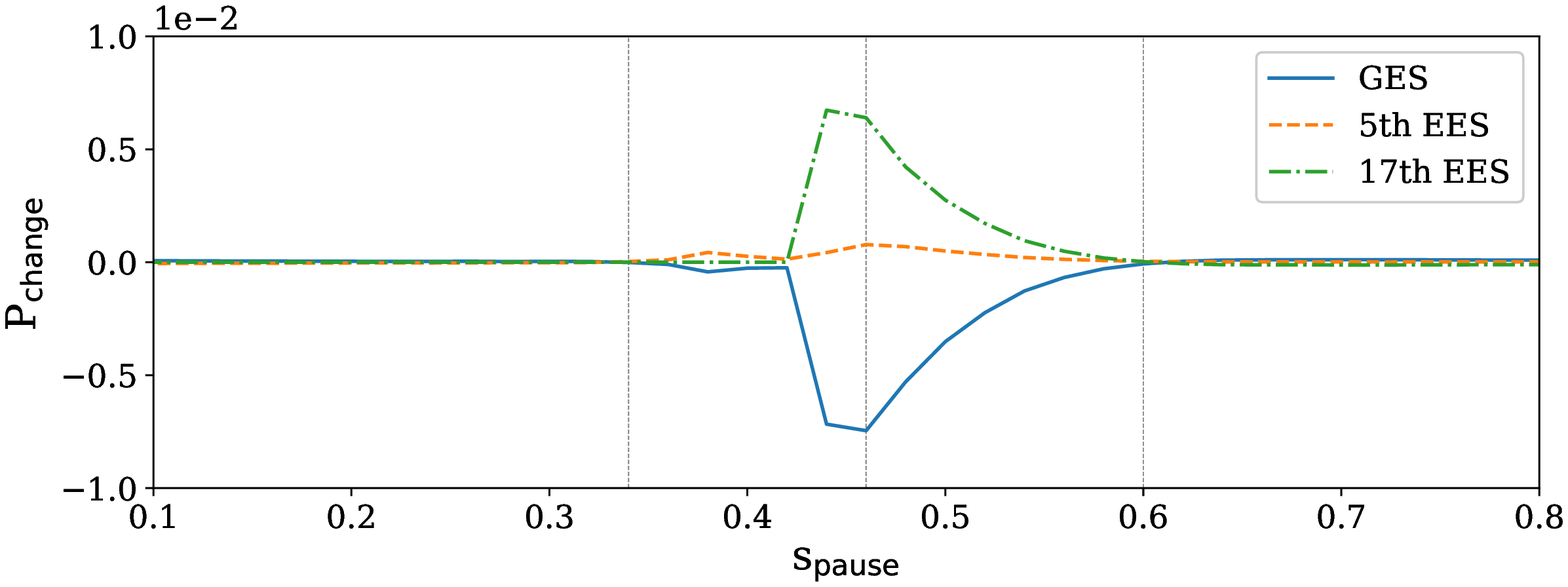}%
    \caption{
        \label{fig_prob_change}
        (Color online) (a) Changes in instantaneous eigenstate probabilities during anneal pause for $\tau_\mathrm{anneal} = \tau_\mathrm{pause} = 1$ and $s_\mathrm{pause} = 0.46$.
        Probability changes for the first 100 eigenstates are plotted.
        (b) Changes in probabilities of the instantaneous ground eigenstate and the fifth and 17th excited eigenstates versus anneal pause parameter $s_\mathrm{pause}$.
        GES, EES: instantaneous ground and excited eigenstate
    }
\end{figure}

\begin{figure}[thb]
    \center
    (a) \includegraphics[width=85mm,clip]{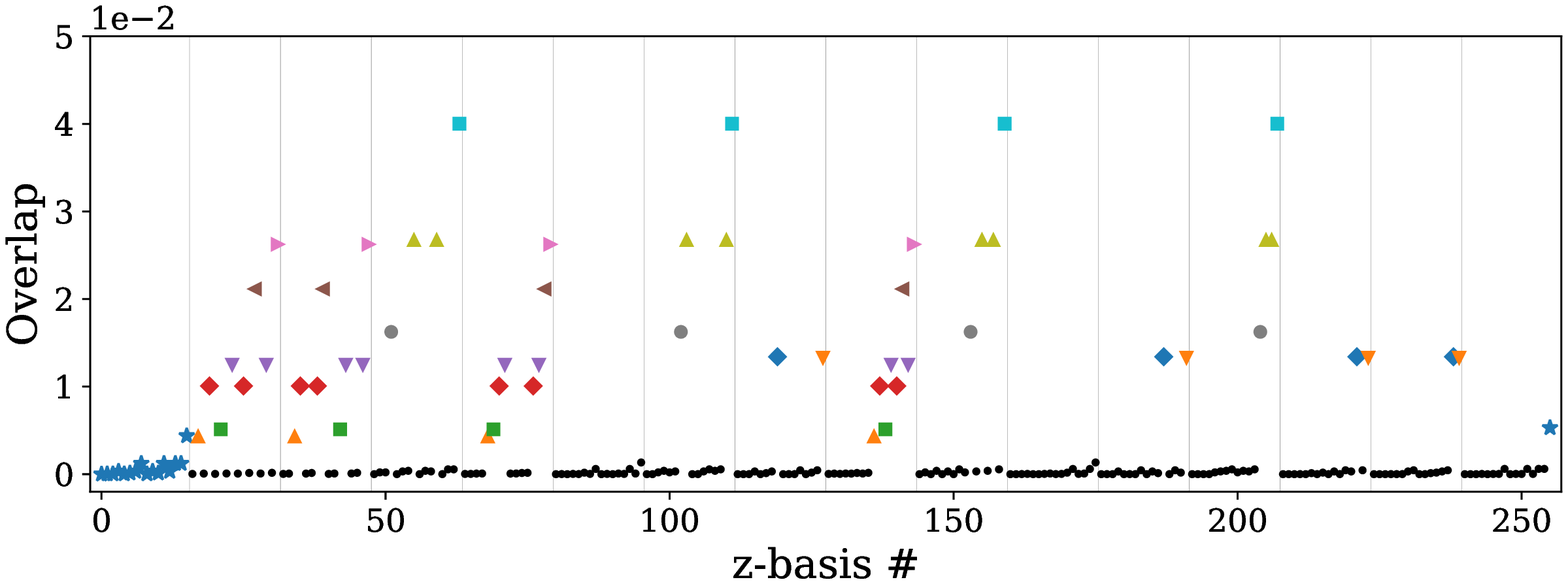} \\
    (b) \includegraphics[width=85mm,clip]{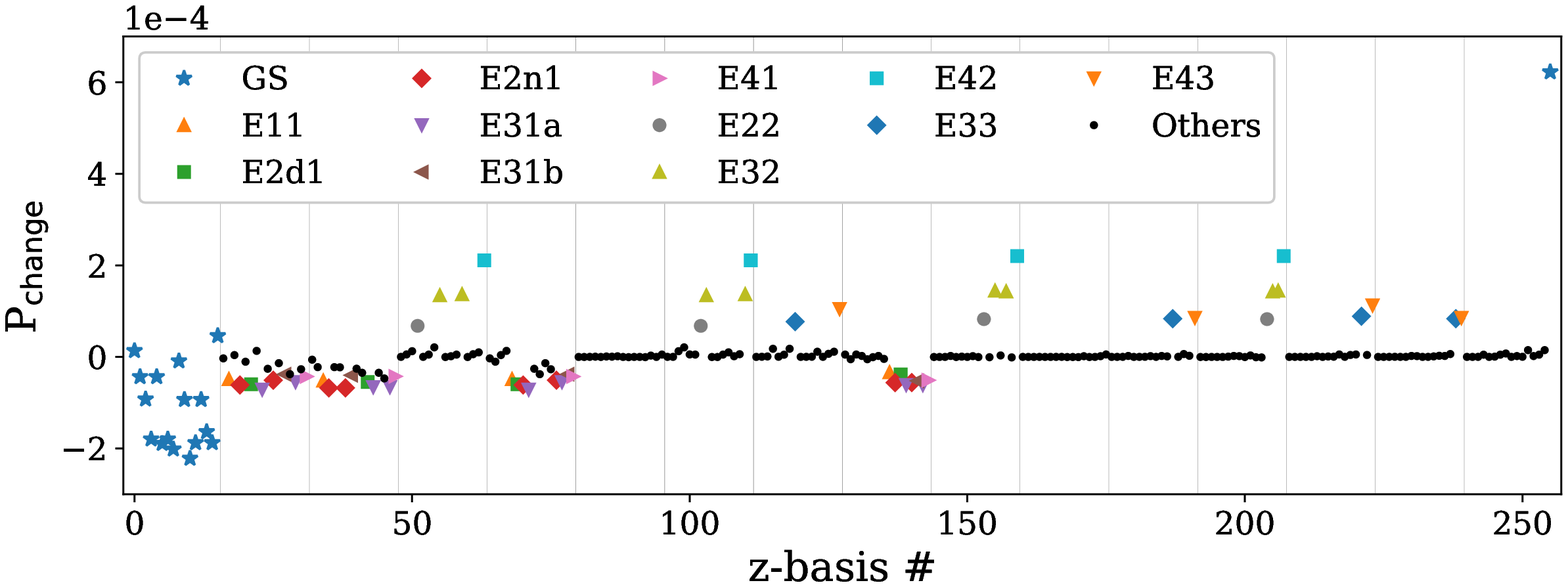}%
    \caption{
        \label{fig_zbasis}
        (Color online) (a) Overlaps between the 17th excited eigenstate and the $z$-basis at $s = 0.46$.
        (b) Probability changes during anneal pause as functions of the $z$-basis for $\tau_\mathrm{anneal} = \tau_\mathrm{pause} = 1$ and $s_\mathrm{pause} = 0.46$.
        The first 16 and the last points represent the clustered and the isolated ground states of $\mathcal{H}_c$ respectively.
        All plots share the same legend.
        GS: classical ground state.
    }
\end{figure}

During the anneal pause, the quantum dynamics with a time-independent Hamiltonian do not essentially drive the system.
Therefore, the probability should be transferred from the clustered ground states to the isolated state by the thermodynamic effects dominantly, which increases $P_s / P_c$.
We investigate the dynamics in instantaneous eigenstates (Fig~\ref{fig_prob_change}) to narrow down excited states involved in the dynamics, then focus on $z$-basis states (Fig.~\ref{fig_zbasis}) to understand probability transfer among the isolated and clustered ground states.

Figure~\ref{fig_prob_change}(a) shows how the simulated wave-function changes during the anneal pause at $s_\mathrm{pause}$ = 0.46 in terms of overlaps against the instantaneous eigenstates.
In Fig.~\ref{fig_prob_change}(a), changes in probability defined as difference of squared overlaps $|\braket{a|\psi(t_2)}|^2 - |\braket{a|\psi(t_1)}|^2$ ($\{\ket{a}\}$: instantaneous eigenstates, $t_1$ and $t_2$: time of pause start and end) are plotted.
The probability of the instantaneous ground state (the eigenstate with the lowest eigenvalue) decreased whereas those of the instantaneous fifth and 17th excited eigenstates increased.
The same phenomena, probability change in these eigenstates, are commonly observed in the region 0.34--0.6, where $P_s / P_c$ ratio increased (see Fig.~\ref{fig_prob_change}(b)).
This observation implies these eigenstates have an essential role in the anneal pause period.
In addition, at the end of the annealing, the fifth excited eigenstate was identified as the isolated state, as noted in an earlier study \cite{Boixo2013} whereas the 17th excited eigenstate was not well characterized.

To characterize the 17th excited eigenstate, the overlaps between the instantaneous 17th excited eigenstates at $s$ = 0.46 and the $z$-basis states in Fig.~\ref{fig_zbasis}(a).
This plot helps to identify essential spin configurations in $\sigma^z$ representation including the isolated and clustered ground states of $\mathcal{H}_c$.
The $z$-basis states highlighted by different shapes and colors have relatively large overlaps, and they are potentially involved in the probability transfer among the isolated and clustered ground states.
While Fig.~\ref{fig_zbasis}(a) is a static analysis, Fig.~\ref{fig_zbasis}(b) is a dynamic analysis and depicts the probability changes in the $z$-basis states during the anneal pause in the numerical simulation.
Highlighted points in both figures from static and dynamic analyses are consistently non-zero and relatively large compared to the non-highlighted points (black dots).
This consistency supports that these figures contain essential information to characterize the 17th excited eigenstate.

\begin{figure}[thb]
    \center
    \includegraphics[width=65mm,clip]{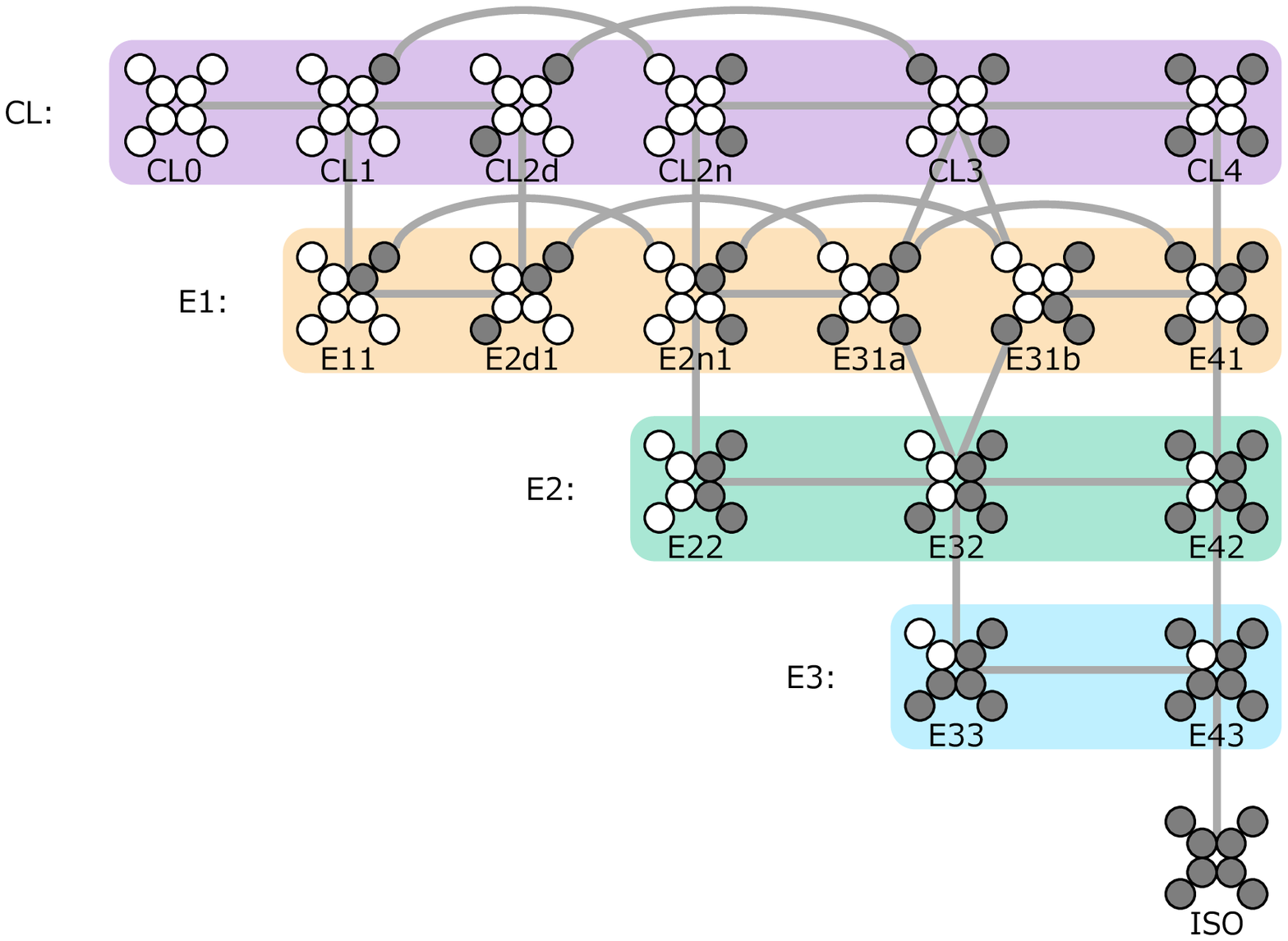}%
    \caption{
        \label{fig_path}
        (Color online) Clustered ground states (top row), isolated ground state (bottom row), and intermediate states (second, third, and fourth rows) involved in probability transfer from the top to the bottom states.
        Connections represent the single-spin flips between the states.
        Open and closed circles represent up and down spins, respectively.
    }
\end{figure}

The $z$-basis states highlighted in Fig.~\ref{fig_zbasis} are located in the path from the clustered to isolated ground states in Fig.~\ref{fig_path}.
The intermediate states can be grouped by the magnetization of their core spins.
Groups with core-spin magnetizations of 2, 1, 0, -1, and -2 are labeled as CL, E1, E2, E3, and ISO, respectively.

\begin{figure}[thb]
    \center
    (a) \includegraphics[width=85mm,clip]{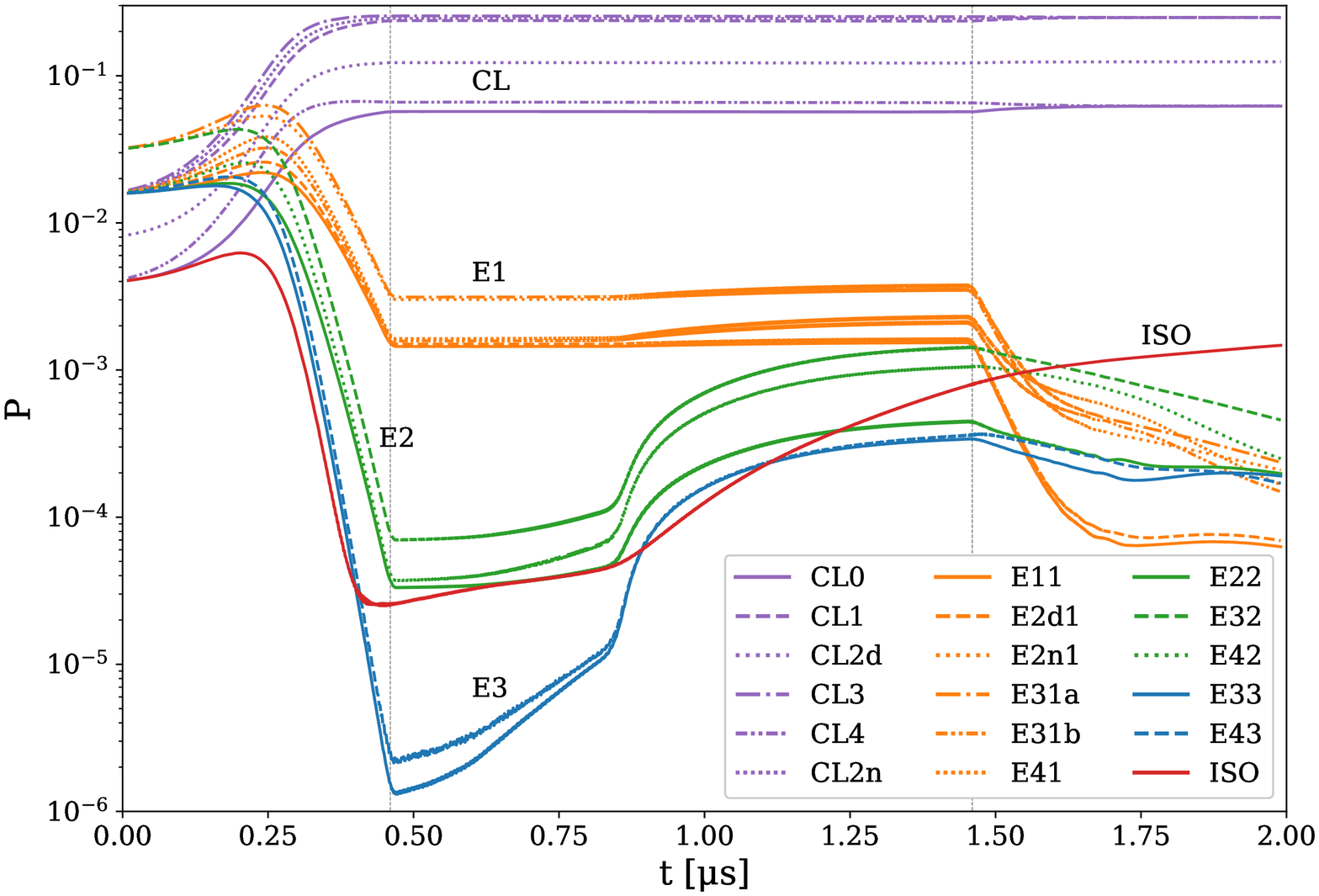} \\
    (b) \includegraphics[width=85mm,clip]{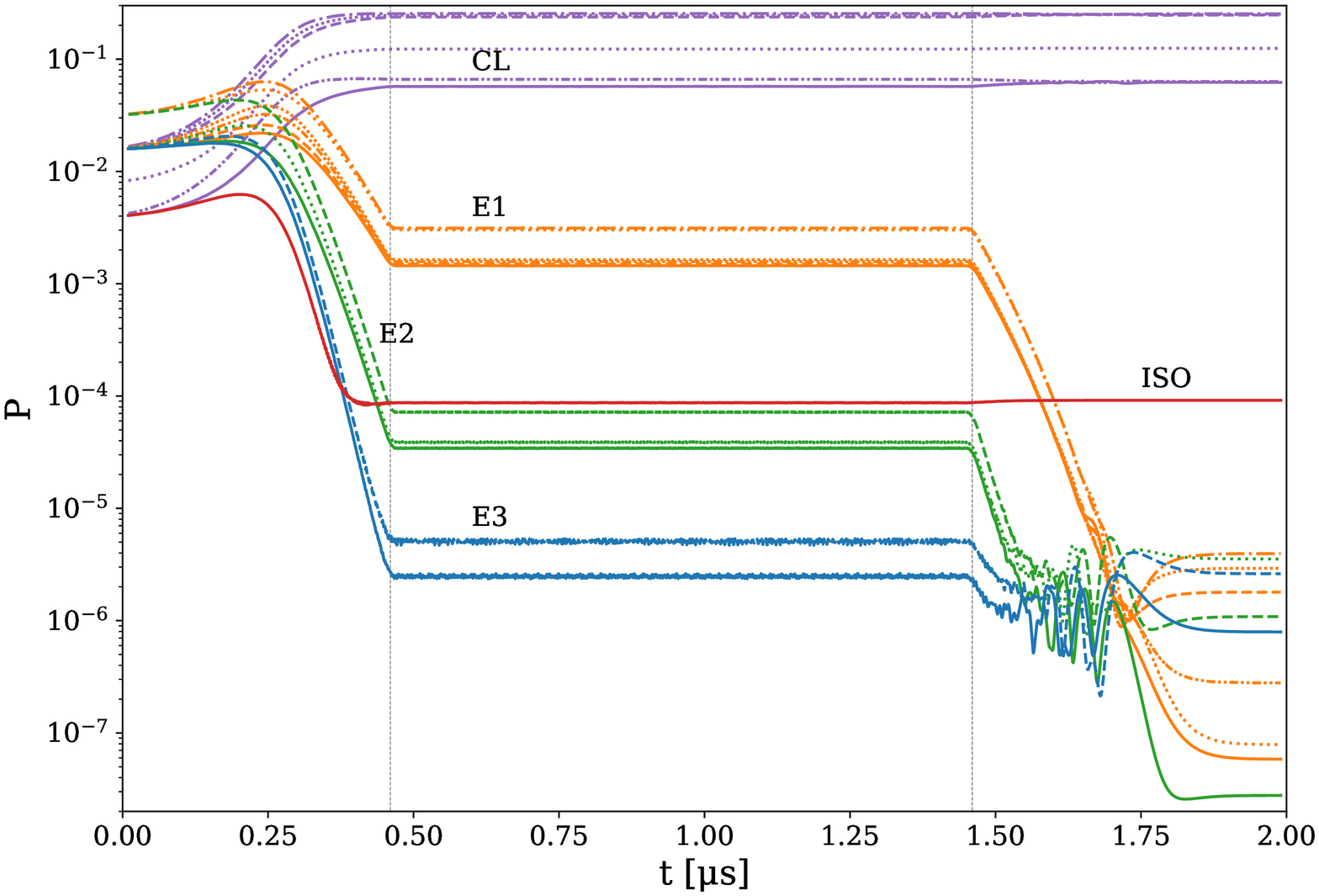}%
    \caption{
        \label{fig_prob}
        (Color online) Temporal changes in probabilities at $s_\mathrm{pause} = 0.46$ and $\tau_\mathrm{anneal} = \tau_\mathrm{pause} = 1$ $\mu$s (a) with energy dissipation ($\alpha = 0.0045$, open quantum dynamics) and (b) without energy dissipation ($\alpha = 0$, closed quantum dynamics).
        As the probability oscillates in the non-eigenbasis, the curves were smoothened by a moving average of 20 ns.
        All plots share the same legend.
    }
\end{figure}

The temporal probability behaviors of these states in the open ($\alpha$ = 0.0045) and closed ($\alpha$ = 0) quantum systems are displayed in panels (a) and (b) of Fig.~\ref{fig_prob}, respectively.
During the annealing period from $t$ = 0.46 to 1.46 $\mu$s, the wave function varied in the open system but remained static (except for the rotating phases) in the closed system.
Referring to the single-spin flip chain shown in Fig.~\ref{fig_path}, the E1 group had a relatively large probability at the start of the anneal pause because its members were directly connected to the clustered ground states.
The lower probabilities of groups E2 and E3 reflect their distances from the clustered ground state.
During the first half of the anneal pause period, the slope of the probability versus time plot was larger in E3 than in the E2 and E1 groups, supporting a probability transfer toward the isolated ground state.

\section{Summary and Discussion}
\label{sec_summary}

Quantum annealers, such as the D-Wave machine, and other quantum devices are operated at the finite temperature and are coupled (albeit weakly) with the environment.
By understanding such quantum devices as open quantum systems, we can expect to improve the control of systems.

The anneal pause function implemented in the D-Wave 2000Q can probe the thermodynamic effects in open quantum systems.
We investigated how the anneal pause modulates the quantum-signature model.
The $P_s / P_c$ ratio in the model reflects the balance between quantum mechanics and thermodynamics.
From the experimental results in Fig.~\ref{fig_experiments}, we found dependences on both the total time of the dynamics and the strength of the transverse field during the anneal pause.
The total time dependence can be explained by accumulative thermodynamic intervention from the environment such that long (slow) dynamics enhance the thermodynamic signature.
The dependence on the strength of the transverse field during the anneal pause is nontrivial and is not well explained by the energy gap structure.
To address these difficulties, we must magnify the phenomenon at the wave function scale in a quantum simulation analysis.

Performing a fine-grained analysis of the system, we employed a recently developed method for open quantum systems\cite{Kadowaki2018}.
This method introduces energy dissipation by interpolating between the quantum dynamics and the thermodynamics.
Although the method provides a numerical interpolation protocol, an accurate analysis requires an analytical form of the nonlinear differential equations that merge the whole process of the protocol.
In the present paper, we established a method for deriving the nonlinear differential equations of multi-spin systems, supported by a symbolic computation program.

By properly selecting the parameters, temperature $T$ and coupling $\alpha$, in the dynamics of the quantum-signature model, we reproduced the two fundamental features observed in the experiment.
The temperature was selected to approximate the operational temperature of the real quantum annealer whereas the coupling strength was fitted to the experimental results.
Another potential parameter, which was not considered in this study, is the timescale difference between the quantum and thermodynamic phenomena.
This parameter was omitted to reduce a large number of experiments and simulations when fitting multiple parameters simultaneously, but must be considered in a quantitative comparison.

From the simulation results, we could extract detailed information on the system dynamics.
The analyses revealed the essential path driven by the 17th excited eigenstate, which transfers probability from the clustered ground states to the isolated ground state (the thermalization mechanism in this model).
Owing to the many large overlaps between the excited eigenstate and the 11 states in the $z$-basis, a chain of single spin-flips establishes between the clustered and isolated ground states.
This phenomenon informs us that thermalization in a quantum system can help to retrieve the classical ground states than pure quantum dynamics.
Mitigating unfair sampling can improve statistical machine learning~\cite{Korenkevych2016}.

For this purpose, we do not have an analytical method to select the best $s_\mathrm{pause}$ to control $P_s / P_c \sim 1$ at this point.
Although we need to conduct a grid search on $s_\mathrm{pause}$, we found that the $P_s / P_c$ ratio can be improved more than one digit by combining longer annealing period and appropriate anneal pause setting (see the experimental results in Fig.~\ref{fig_experiments}).
Further investigation can provide a better method to select $s_\mathrm{pause}$.
In addition to the previous findings of thermal relaxation after anticrossing\cite{Amin2008,Venuti2017,Dickson2013}, we found another role of thermal relaxation that transfers probabilities among ground states, then mitigates imbalance of probabilities.
This is an example of how quantum and thermal effects work synergistically in quantum annealing.

For feasible open-quantum simulations, the method solves the nonlinear differential equations of $2^n$ complex variables originating from $n$-qubits without introducing the thermal bath variables.
Relaxation of the system by the thermal bath is incorporated directly to the differential equations of the system through the interpolation with master equation parametrized by $\alpha$.
This construct reduces the computational cost by removing the Hilbert space expansion for the thermal bath.
Conversely, nonlinear differential equations are more computationally costly to construct and solve than linear ones.
In addition, which systems can be resolved by this method is only partially answered in the present and the previous studies.
We demonstrated the qualitative usefulness of this method by reproducing the trivial and nontrivial phenomena observed in open quantum systems.
In the current formulation, the system--bath coupling is assumed as $\sigma^z$ interaction as our master equation is formulated by $\sigma^z$.
Beyond this assumption, expansion of the method will be required.

In future work, the potential of the fair-sampling feature should be confirmed in experiments and theoretical analyses using open-quantum simulations.
A former study of different models found that the classical ground state tends to accumulate more probabilities when the number of free spins is increased\cite{Matsuda2009}.
The same conclusion was reached in our current analysis.
Thus, we expect that fair sampling will emerge in the models\cite{Matsuda2009} and in more general cases.
The biased sampling problem and possible performance improvements using thermal fluctuations have also been discussed\cite{Mandra2017}.
The authors suggested hybrid architectures that encourage thermal fluctuations, as also supported by our findings.

Our method is not only limited to simulate quantum annealers but also can be applied to enhance the understanding of quantum devices, such as noisy intermediate-scale quantum devices.
Such further applications would reveal the usefulness and limitations of the method.

\section*{Acknowledgment}
We are grateful to S. Tanaka, M. J. Miyama, H. Irie and S. Okada and D-Wave support team for their helpful discussions and suggestions.
We are also grateful to the reviewer for pointing out the relation between our proposed method and stochastic Schr\"odinger equation.
Part of the work is financially supported by MEXT KAKENHI Grant No. 15H03699 and 16H04382, and by ImPACT and JST START.

\bibliography{refs}

\appendix
\section{Derivation for a two-level system}
We demonstrate to derive differential equations of ID for a two-level system.
The system consists of a spin interacting with the virtical field $\mathcal{H}_c = -h \sigma^z$ and the transverse field $\mathcal{H}_q = -\Gamma \sigma^x$.
Thus, a matrix representation of the Hamiltonian is,
\begin{equation}
  \mathcal{H} = \mathcal{H}_c + \mathcal{H}_q = \left(
  \begin{matrix}
    -h & -\Gamma \\
    -\Gamma & h
  \end{matrix}
  \right)
\end{equation}
Using the diagonal part of the Hamiltonian, we have the following transition rate matrix,
\begin{equation}
  \mathcal{L} = \frac{1}{e^{\beta h} + e^{-\beta h}} \left(
  \begin{matrix}
    -e^{-\beta h} & e^{\beta h} \\
    e^{-\beta h} & -e^{\beta h}
  \end{matrix}
  \right) ,
\end{equation}
where $\beta$ is the inverse temperature.
For simplicity, we take a zero temperature limit, $\beta \to \infty$.
Then, the matrix is
\begin{equation}
  \mathcal{L} = \left(
  \begin{matrix}
    0 & 1 \\
    0 & -1
  \end{matrix}
  \right) .
\end{equation}
A wave function of Schr\"odinger eqation and a probability distribution of master equation are two components, and we assume that they can be expressed using the same functions, $u(t)$ and $d(t)$,
\begin{equation}
  \ket{\psi(t)} = \left(
  \begin{matrix}
    u(t) \\
    d(t)
  \end{matrix}
  \right) \quad \mbox{and} \quad \bm{P}(t) = \left(
  \begin{matrix}
    |u(t)|^2 \\
    |d(t)|^2
  \end{matrix}
  \right)
\end{equation}
The following equations are obtained by substituting them for Eq.~(\ref{eq_new_state}),
\begin{align}
  u(t+dt) = & \ \sqrt{(1-\alpha) \big|u(t)+\big[ihu(t)+i\Gamma d(t)\big]dt\big|^2 + \alpha\Big(|u(t)|^2+|d(t)|^2dt\Big)} \nonumber \\
  & \times \frac{u(t)+\big[ihu(t)+i\Gamma d(t)\big]dt}{\big|u(t)+\big[ihu(t)+i\Gamma d(t)\big]dt\big|} \\
  d(t+dt) = & \ \sqrt{(1-\alpha) \big|d(t)+\big[i\Gamma u(t)-ihd(t)\big]dt\big|^2 + \alpha\Big(|d(t)|^2-|d(t)|^2dt\Big)} \nonumber \\
  & \times \frac{d(t)+\big[i\Gamma u(t)-ihd(t)\big]dt}{\big|d(t)+\big[i\Gamma u(t)-ihd(t)\big]dt\big|}
\end{align}
By using series expansion for $dt$ we have difference equations,
\begin{align}
  u(t+dt) - u(t) & = \bigg\{ ihu(t) + i\Gamma d(t) + \frac{\alpha}{2}\Big[\frac{|d(t)|^2}{u^*(t)} - i\Gamma\Big(d(t) - d^*(t)\frac{u(t)}{u^*(t)}\Big)\Big] \bigg\} dt + O(dt^2) \\
%  u(t+dt) - u(t) & = \bigg\{ ihu(t) + i\Gamma d(t) + \frac{\alpha}{2}\Big[\frac{1}{u^*(t)} - u(t) - i\Gamma\Big(d(t) - d^*(t)\frac{u(t)}{u^*(t)}\Big)\Big] \bigg\} dt + O(dt^2) \\
  d(t+dt) - d(t) & = \bigg\{ i\Gamma u(t) - ihd(t) + \frac{\alpha}{2}\Big[- d(t) - i\Gamma\Big(u(t) - u^*(t)\frac{d(t)}{d^*(t)}\Big)\Big] \bigg\} dt + O(dt^2)
%  d(t+dt) - d(t) & = \bigg\{ i\Gamma u(t) - ihd(t) - \frac{\alpha}{2}\Big[d(t) - i\Gamma\Big(u^*(t)\frac{d(t)}{d^*(t)} - u(t)\Big)\Big] \bigg\} dt + O(dt^2)
\end{align}
Finally differential equations of the two level system are derived as,
\begin{align}
  \label{eq_tl_up}
  \frac{d}{dt} u(t) & = ihu(t) + i\Gamma d(t) + \frac{\alpha}{2}\Big[\frac{|d(t)|^2}{u^*(t)} - i\Gamma\Big(d(t) - d^*(t)\frac{u(t)}{u^*(t)}\Big)\Big] \\
%  \frac{d}{dt} u(t) & = ihu(t) + i\Gamma d(t) + \frac{\alpha}{2}\Big[\frac{1}{u^*(t)} - u(t) - i\Gamma\Big(d(t) - d^*(t)\frac{u(t)}{u^*(t)}\Big)\Big] \\
  \label{eq_tl_down}
  \frac{d}{dt} d(t) & = i\Gamma u(t) - ihd(t) + \frac{\alpha}{2}\Big[- d(t) - i\Gamma\Big(u(t) - u^*(t)\frac{d(t)}{d^*(t)}\Big)\Big]
%  \frac{d}{dt} d(t) & = i\Gamma u(t) - ihd(t) - \frac{\alpha}{2}\Big[d(t) - i\Gamma\Big(u^*(t)\frac{d(t)}{d^*(t)} - u(t)\Big)\Big]
\end{align}
One can confirm that these equations are equivalent to the density matrix representation previously reported\cite{Kadowaki2018},
\begin{align}
  \frac{d}{dt}\big[u(t)u^*(t)\big] = \frac{d}{dt}\rho_{11} & = -i\Gamma(1-\alpha)(\rho_{12}-\rho_{21}) + \alpha\rho_{22} , \\
  \frac{d}{dt}\big[u(t)d^*(t)\big] = \frac{d}{dt}\rho_{12} & = -i\Gamma\Bigg\{1-\frac{1}{2}\bigg[1-\Big(\frac{\rho_{12}}{|\rho_{12}|}\Big)^2\bigg]\Bigg\}(\rho_{11}-\rho_{22}) - \bigg(\frac{\alpha(\rho_{11}-\rho_{22})}{2\rho_{11}}-2ih\bigg)\rho_{12} .
\end{align}

\section{Generation of C code for simulation}
This Mathematica code generates C code to calculate RHS of the interpolated dynamics for the two-level system in Appendix A. The eight-spin quantum-signature model can be calculated by defining $H$, $L$, $\psi$, and $P$ in the same manner.
\begin{align*}
    & (* \ \mbox{Definition of the system} \ *) \\
    & H = \{\{-h, -\Gamma\}, \{-\Gamma, h\}\} \\
    & L = \{\{0, 1\}, \{0, -1\}\} \\
    & \psi = \{u_x+I \ u_y, \ d_x+I \ d_y\} \\
    & P = \{u_x^2+u_y^2, \ d_x^2+d_y^2\} \\
    & (* \ \mbox{Calculate derivatives} \ *) \\
	& d\psi = -I \ H . \psi \ dt \\
    & dP = L.P \ dt \\
    & \psi_{next} = \psi + d\psi \\
	& P_{next} = P + dP \\
    & (* \ \mbox{Calculate RHS} \ *) \\
    & \psi 2_{next} = \mbox{ComplexExpand}[\mbox{Conjugate}[\psi_{next}] \psi_{next}] \\
    & r = \sqrt{(1 - \alpha) \ \psi 2_{next} + \alpha \ P_{next}} \\
    & ph = \psi_{next} / \sqrt{\psi 2_{next}} \\
    & rhs = \mbox{Coefficient}[\mbox{Series}[r \ ph - \psi, \{dt, 0, 1\}], dt] \\
    & (* \ \mbox{Output RHS in C code} \ *) \\
    & \mbox{Write}[\mbox{OpenWrite}[\mbox{"RHS.c"}], \mbox{CForm}[rhs]] \\
\end{align*}

\end{document}